\documentclass[11pt]{article}

\usepackage[margin=1in]{geometry}
\usepackage[T1]{fontenc}
\usepackage{lmodern}
\usepackage{microtype}
\usepackage{setspace}
\usepackage{graphicx}
\usepackage{booktabs}
\usepackage{longtable}
\usepackage{threeparttable}
\usepackage{threeparttablex}
\usepackage{array}
\usepackage{caption}
\usepackage{subcaption}
\usepackage{float}

\usepackage{multirow}

\usepackage{amsmath}
\usepackage{amssymb}
\usepackage{mathtools}
\usepackage{bbm}

\usepackage{enumitem}
\setlist[itemize]{leftmargin=*,topsep=2pt,itemsep=2pt}
\setlist[enumerate]{leftmargin=*,topsep=2pt,itemsep=2pt}

\usepackage[hyphens]{url}
\usepackage[colorlinks=true,linkcolor=blue!50!black,citecolor=blue!50!black,urlcolor=blue!50!black,breaklinks=true]{hyperref}
\usepackage{cleveref}
\usepackage{xcolor}

\usepackage[numbers,sort&compress]{natbib}

\providecommand{\doi}[1]{}

\usepackage{fancyhdr}
\graphicspath{{figures/}{../../4.results/figures/}}

\crefname{section}{Section}{Sections}
\crefname{figure}{Figure}{Figures}
\crefname{table}{Table}{Tables}
\crefname{equation}{Equation}{Equations}

\usepackage{authblk}

\title{\textbf{An economic geography dataset of U.S. skill specialization, relatedness, and complexity}}

\author[1]{Anthony Howell\thanks{Corresponding author: \href{mailto:ajhowel5@asu.edu}{Anthony.Howell@asu.edu}}}
\author[1]{Maryann Feldman}
\author[2]{Lauren Lanahan}
\author[3]{Nikhil Kalathil}
\author[4]{Evan Johnson}

\affil[1]{Arizona State University}
\affil[2]{University of Oregon}
\affil[3]{Carnegie Mellon University}
\affil[4]{University of North Carolina at Chapel Hill}

\date{\today}

\begin{document}
\maketitle

\begin{abstract}
\noindent We release a new dataset of U.S.\ skill specialization, relatedness, and complexity, derived from 433.6 million Lightcast job postings between 2010 and 2024. The panel covers 3,194 counties across 15 years and reports 201 variables that describe the volume of job postings (e.g., labor demand), the modality and nature of work (e.g., remote share, internship share), and the structure of employer skill demand by category (e.g., specialized, software, and common).  We develop a suite of economic geography variables:  skill-based measures of county specialization, relatedness, diversity, complexity, and dynamics. These measures are further decomposed by employer entity type (corporate, university, government, and federal lab), along with entity-pair measures of alignment, overlap, and directional skill gaps. An accompanying interactive dashboard supports both academic research and applied use, with features including spatiotemporal visualization, county rankings and trends, pairwise county comparisons, and individual county profiles.

\end{abstract}

\noindent\textbf{Keywords:} economic geography, skill demand, skill relatedness, skill complexity, skill dynamics.

\vspace{1em}

\newpage
\clearpage

\section{Background \& Summary}
\label{sec:background}

The skills that employers seek vary across cities, regions, and labor-market areas in ways that shape regional productivity, wage inequality, and innovation \citep{moretti2012new, diamond2016determinants, autor2019work, balland2020complex}. Decades of research in economic geography, labor economics, and regional science have made clear that aggregate national statistics conceal defining features of modern labor markets. Places differ in the kinds of work employers need and their skill requirements, not merely in how much \citep{bartik2020bartik}. This variation is not random. It is structured by what each place already does well, and this path-dependent structure both constrains and enables what each place can become.

Economic geography has developed an increasingly unified analytical framework for measuring this structure. The framework rests on a bipartite mapping between locations and the activities they specialize in \citep{neffke2011how, boschma2017relatedness, hidalgo2018principle}, and on a common family of measures derived from that mapping: specialization, relatedness, diversity, and complexity \citep{balland2022newparadigm}. These measures are unit-agnostic, applying across organizational and geographic units of analysis: the same mathematical objects have been used for export products \citep{hidalgo2007product}, industries \citep{neffke2011how}, patent technologies \citep{kogler2013mapping}, occupations \citep{alabdulkareem2018unpacking}, and skills \citep{neffke2013skill}. At each instantiation, they predict regional wages, growth, resilience, firm entry, innovation, and worker mobility in ways that aggregate employment counts and industry classifications cannot \citep{diodato2018industries}.

Skills sit at the foundation of this measurement stack. They are the
underlying capability units that determine which trajectories a region
can plausibly enter. Skill-based measures tend to outperform
industry-based measures for predicting firm diversification, worker
mobility, and the formation of new specializations \citep{neffke2013skill}.
They also evolve on a much shorter timescale than the industry or 
technology codes that have historically anchored regional analysis: 
technological change, new occupations, and shifting employer requirements 
rewrite the skill landscape on a yearly horizon \citep{deming2020earnings}, while industry and 
technology classifications turn over on a multi-year cycle and 
necessarily lag the emergence of new activities, capabilities, and 
occupations.

Yet most empirical work to date has had to rely on indirect proxies for skill demand, such as occupational mixes, O*NET task content imputed from job titles, or survey-based measures coarse in time and space \citep{alabdulkareem2018unpacking}. Direct, fine-grained measures of what employers actually ask for at the county-year level remain unavailable in publicly released form.  This limits empirical work on regional skill ecosystems, the spatial diffusion of capabilities, and the design of place-based policy.

A brief inventory of existing public data sources clarifies the gap. The U.S. Bureau of Labor Statistics (BLS) Occupational Employment and Wage Statistics reports counts and wages by occupation and metropolitan area but contains no measure of skill content within occupation. The BLS Job Openings and Labor Turnover Survey measures vacancy counts at the national and state level only, not by occupation or skill and not at sub-state geography. 

The O*NET database describes the typical skill content of occupations but is a cross-sectional, expert-coded snapshot that does not vary across regions and updates infrequently. Census Bureau products including County Business Patterns and the Quarterly Workforce Indicators provide county-industry employment statistics but carry no skill information. Similarly, the linked Longitudinal Employer-Household Dynamics captures worker flows but contains no skill content.

Vendor-curated job-posting datasets have partially filled this gap in research practice. Data from LinkedIn, Indeed, and Lightcast (formerly Burning Glass) underpin a growing literature on skill demand and recessions \citep{deming2018skill,hershbein2018recessions}, the rising importance of social and cognitive skills \citep{deming2017growing}, AI exposure and adoption \citep{acemoglu2022ai,alekseeva2021demand,babina2024artificial}, and the geography of AI work \citep{felten2021occupational}. The underlying micro data are licensed and require institutional access. Publicly released aggregate products derived from these data are rare. A notable exception is the work of Azar, Marinescu, and colleagues, who use online vacancy micro data to construct and release labor-market concentration measures, supporting new work on employer monopsony power \citep{azar2022labor}.

Building on this precedent, we release the most comprehensive geography of 
skills panel to date. Spanning 2010--2024 and built from 433.6 million 
Lightcast job postings in the contiguous United States, Alaska, and 
Hawaii, our county-year panel covers 3{,}194 counties and 15 years, 
producing 47{,}891 county-year observations and 201 variables per 
observation. The panel operationalizes the full economic geography 
measurement stack (specialization, diversity, and complexity) through 
direct measures of the structure, composition, and dynamics of 
employer-based skill demand, and extends that stack with entity-type 
decomposition and pairwise similarity measures absent from existing public datasets.  An accompanying interactive 
dashboard provides dynamic map visualizations of key indicators, 
bivariate relationships, and in-depth county-specific profiles.

The variables characterize local labor and skill demand along three main dimensions. \textit{Who is hiring}: total postings, total skill-mention counts by skill type, and posting counts by employer entity type, decomposed across four types (corporate, university, federal lab, and government). \textit{The nature of work}: posting counts and shares by modality (remote, on-site) plus the count of internship-flagged postings. \textit{The composition and structure of skill demand}: skill-portfolio composition, specialization, relatedness, diversity, complexity, network position, and year-over-year dynamics.

Operationally, we split the variables into two parts. A 37-variable core set
(groups A--H) collects the aggregate county-year measures. A 164-variable 
entity-decomposed extension adds entity-specific measures of each entity 
type's own specialization breadth and year-over-year dynamics (groups I 
and J) and entity-pair measures of alignment, overlap, and directional 
skill gaps between any two entity types within the same county-year 
(group K).


The panel supports research questions on regional capability formation that existing public data could not easily answer. Skill density and churning measure the local branching process directly, rather than through industry or patent proxies \citep{boschma2017relatedness,neffke2011how}. County-level Economic Complexity Index (ECI) and fitness series enable analogous tests to the cross-country and patent-based complexity literature \citep{hidalgo2009building,balland2017geography,mewes2022technological}. Cosine distance and net Revealed Comparative Advantage (RCA) churn provide intensive- and extensive-margin views of demand reorganization after shocks, informing work on regional resilience \citep{martin2015notion,diodato2015resilience}.

Skill-based agglomeration measures complement industry-based co-location indices \citep{ellison2010coagglomeration,diodato2018industries,howell2017marshallian,howell2023agglomeration}.  The panel could be used to explore whether the CHIPS Act, IRA, and Tech Hubs designations build on or diverge from existing county skill profiles \citep{balland2019smart,juhasz2024new}. The entity-decomposed extension also enables comparisons of public-, academic-, and private-sector skill demand within and across counties \citep{neffke2013skill,diodato2018industries}.


\section{Methods}
\label{sec:methods}

This section describes the transformation pipeline from the raw Lightcast job-posting micro data (929 GB across 22,967 gzipped CSV shards, 433.6 million postings, 2010--2024) to the released county-year panel. The pipeline is organized into two phases. Phase A streams through the raw postings once, extracts skill mentions and posting characteristics, classifies each posting's employer into one of four entity types (Corporate, University, Government, and Federal Lab), and aggregates to per-year checkpoint tables keyed by county, skill, and employer type. Phase B relies on the Phase A outputs to compute the derived economic-geography measures that constitute the released panel.

\subsection{Data source}
\label{sec:methods:source}

The raw input is the Lightcast (formerly Burning Glass Technologies, formerly Emsi Burning Glass) Main job-posting file for calendar years 2010 through 2024. Each row represents one job posting parsed from a U.S.~employer's online advertisement. Lightcast aggregates postings from over 50{,}000 unique online sources and de-duplicates near-identical reposts. The parser produces 134 structured fields covering job title, employer name, geography (state, MSA, city, county), salary when reported, and work-mode indicators (remote, hybrid, on-site).

Skills in the Lightcast taxonomy are organized into three categories: specialized skills (occupation-specific technical skills, e.g.~``machine learning''), software skills (named software products and platforms, e.g.~``AutoCAD''), and common skills (transferable soft skills, e.g.~``customer service''). Across the three categories, 29{,}256 distinct skill names appear in the data. 

\subsection{Phase A: streaming extraction and aggregation}
\label{sec:methods:phasea}

Phase A scans the raw postings once, year by year, and produces three per-year output tables.

\textbf{Geography and time.}\quad We date postings by their \texttt{posted} field and locate them by 5-digit county FIPS code. We retain and flag state-level FIPS codes (ending 999); they account for less than 1 percent of postings overall but are over-represented among remote postings and the postings of large multi-location employers.

\textbf{Skill parsing.}\quad We parse skills from the three Lightcast skill columns (specialized, software, and common) into a long-format mentions table keyed by (\texttt{county}, \texttt{skill\_name}, \texttt{skill\_type}). We keep postings with no listed skills for posting-count purposes; they contribute zero skill rows.

\textbf{Employer-type classification.}\quad We partition employers into four entity types that capture the principal institutional sectors of the U.S.~labor market: government, universities, federal research laboratories, and the private sector. The first three are public or public-serving institutions whose skill demand is of independent policy interest, and the fourth is the private sector that constitutes the bulk of employment. The government type (NAICS sector 92, public administration) collects federal, state, and local government, and the university type (NAICS 6113, colleges, universities, and professional schools) collects degree-granting institutions. The federal lab type collects a small number of federally funded research and development centers (for example, NASA facilities), identified through named-entity matching on employer name.

The fourth type, the private sector, is the residual: every posting not positively identified as government, university, or federal lab, including postings with missing or ambiguous industry information. It loosely captures the incorporated and other private-sector employers that make up the non-institutional remainder. We do not partition the private sector further because industry classification is absent for a substantial share of postings, ranging from roughly 40 percent in the early years of the panel to 26 percent in the later years. At that rate of non-coverage a finer split would be too incomplete to support reliable entity-specific measures.

\textbf{Outputs.}\quad For each year, Phase A produces three tables: (i) \texttt{skill\_counts}, county-skill-year mention counts; (ii) \texttt{employer\_skill}, county-employer-skill-year mention counts; and (iii) \texttt{panel\_stats}, county-year totals by posting characteristic (posting count, skill-mentioned postings, remote mode, internships).

\subsection{Phase B: derived measures}
\label{sec:methods:phaseb}

Phase B computes the economic-geography measures that constitute the released county-year panel from the Phase A tables.

\textbf{Measurement lineage.}\quad Each released measure adapts a 
published construction from the economic complexity and economic geography 
literatures to the county-skill setting. RCA is from \citet{balassa1965trade},
originally for export specialization. The skill-skill relatedness matrix 
and skill density adapt the product-space construction of 
\citet{hidalgo2007product}; the smart-specialization variant of density 
in \citet{balland2019smart} is also reflected.  Skill coherence follows the 
related-variety formulation in \citet{neffke2011how} and \citet{boschma2017relatedness}. 

The ECI implements the method-of-reflections of
\citet{hidalgo2009building}; \citet{mealy2019interpreting} show that ECI
is mathematically equivalent to the second eigenvector of a normalized-cut 
spectral clustering, which qualifies its interpretation as a single 
sophistication ladder and supports a clustering reading alongside the 
ordering reading. Fitness-complexity \citep{tacchella2012new} is a 
non-linear alternative to the linear reflection. \citet{balland2017geography} 
apply ECI-style methods to patent data and provide the closest 
US-data precedent for the present panel. 

\citet{neffke2013skill} develops 
an alternative skill-relatedness construction based on cross-industry 
labor flows; we do not use it directly but note it as the main parallel 
to the co-occurrence construction we adopt. Each of these constructions 
has prior validation in cross-country product data and cross-region 
industry and patent data; we apply them here to U.S.~county-level skill 
data.

\subsubsection{Revealed comparative advantage}

For each year $t$, each county $c$, and each skill $s$, we compute the Balassa RCA \citep{balassa1965trade}:
\begin{equation}
  \mathrm{RCA}_{c,s,t} \;=\; \frac{X_{c,s,t} \,/\, \sum_{s'} X_{c,s',t}}{\sum_{c'} X_{c',s,t} \,/\, \sum_{c'}\sum_{s'} X_{c',s',t}},
  \label{eq:rca}
\end{equation}
where $X_{c,s,t}$ is the number of times skill $s$ is mentioned in county $c$ in year $t$. Values above one indicate that skill $s$ is more concentrated locally than in the national skill pool. RCA is computed for the full Cartesian product of counties and skills in each year; zero-count cells receive $\mathrm{RCA} = 0$. The same quantity is computed within each employer-type slice (corporate, university, federal lab, and government).

\textbf{Ubiquity.}\quad For each skill-year, ubiquity is the number of counties with $\mathrm{RCA}_{c,s,t} > 1$:
$$\mathrm{Ubiquity}_{s,t} = \sum_c \mathbbm{1}[\mathrm{RCA}_{c,s,t} > 1].$$
Ubiquity is the skill-side analog of diversification: low-ubiquity skills are local specializations of few counties; high-ubiquity skills are widespread. The released panel reports the per-county average of ubiquity over the county's $\mathrm{RCA}>1$ skills (\texttt{avg\_ubiquity}).

\subsubsection{Skill-skill relatedness}

We compute the pairwise skill-skill relatedness matrix per year following \citet{hidalgo2007product}. Let $M_{c,s,t} = \mathbbm{1}[\mathrm{RCA}_{c,s,t} > 1]$ be the bipartite indicator of county-skill specialization. The conditional co-occurrence probability is
\begin{equation}
  \phi^{\text{co}}_{s,s',t} \;=\; \frac{\sum_c M_{c,s,t} \, M_{c,s',t}}{\max\!\big(\sum_c M_{c,s,t},\, \sum_c M_{c,s',t}\big)},
\end{equation}
which equals the minimum of $P(s \mid s')$ and $P(s' \mid s)$. The matrix $\Phi_t = [\phi^{\text{co}}_{s,s',t}]$ is sparse (pairs with $\phi < 0.05$ are dropped) and feeds the density, coherence, and centrality calculations below.

\subsubsection{Skill density}

For each county-skill cell, skill density measures how proximate the cell is to the county's existing specialization portfolio \citep{hidalgo2007product,balland2019smart}:
\begin{equation}
  \mathrm{Density}_{c,s,t} \;=\; \frac{\sum_{s' \neq s} \phi^{\text{co}}_{s,s',t} \, M_{c,s',t}}{\sum_{s' \neq s} \phi^{\text{co}}_{s,s',t}}.
\end{equation}
A county-skill pair with high density is one where many of the county's $\mathrm{RCA}>1$ skills are highly related to skill $s$; this predicts entry into $\mathrm{RCA}>1$ on $s$ in subsequent years. Following \citet{balland2019smart}, the released panel reports \texttt{skill\_density} as the per-county-year average of $\mathrm{Density}_{c,s,t}$ taken over skills that the county has \textbf{not} yet specialized in ($s \notin S_{c,t}$). This measures proximity-weighted potential for new specialization, rather than internal cohesion of the existing portfolio (the latter is captured by \texttt{skill\_coherence} below).

\subsubsection{Skill coherence}

For each county-year, coherence is the average pairwise relatedness among the county's $\mathrm{RCA}>1$ skills \citep{neffke2011how,boschma2017relatedness}:
\begin{equation}
  \mathrm{Coherence}_{c,t} \;=\; \frac{1}{|S_{c,t}| (|S_{c,t}| - 1)} \sum_{s \in S_{c,t}} \sum_{s' \in S_{c,t}, s' \neq s} \phi^{\text{co}}_{s,s',t},
\end{equation}
where $S_{c,t} = \{s : \mathrm{RCA}_{c,s,t} > 1\}$. High coherence means a county's specializations form an internally related cluster; low coherence means specializations are scattered across the skill space.

\subsubsection{Economic complexity and fitness}

We compute the ECI following the method-of-reflections framing of \citet{hidalgo2009building}. Letting $k^0_c = \sum_s M_{c,s,t}$ (county diversification) and $k^0_s = \sum_c M_{c,s,t}$ (skill ubiquity), the iteration alternates:
\begin{equation}
  k^{n}_c = \frac{1}{k^0_c} \sum_s M_{c,s,t} \, k^{n-1}_s, \qquad k^{n}_s = \frac{1}{k^0_s} \sum_c M_{c,s,t} \, k^{n-1}_c.
\end{equation}
\citet{mealy2019interpreting} show that the limit of this iteration is the eigenvector associated with the second-largest eigenvalue of the normalized co-occurrence matrix $\tilde{M} = \mathrm{diag}(k^0_c)^{-1} M \, \mathrm{diag}(k^0_s)^{-1} M^{\top}$. We compute ECI directly as that eigenvector, which is numerically exact and avoids the convergence sensitivity of the iterative form. The county ECI is then standardized to mean zero and unit variance within each year. We additionally implement the Tacchella fitness-complexity iteration \citep{tacchella2012new} with 50 iterations as a non-linear alternative; \texttt{fitness} is included in the panel and should be log-transformed or winsorized before use because the iteration is numerically unstable for very-diversified or very-specialized counties.

\subsubsection{Skill-space centrality}

For each county-year, we compute the average network centrality of the county's $\mathrm{RCA}>1$ skills within the year-specific relatedness network $\Phi_t$. We use weighted degree centrality: each skill's centrality is $\sum_{s'} \phi^{\text{co}}_{s,s',t}$, and the county measure is the mean over $s \in S_{c,t}$. High centrality means the county specializes in core, highly-connected skills; low centrality means peripheral specializations.

\subsubsection{Dynamics}

We measure year-over-year structural change with two complementary measures.

\textbf{RCA churning} (extensive margin):
\begin{align*}
  \mathrm{Entries}_{c,t} &= |\{s : \mathrm{RCA}_{c,s,t} > 1 \text{ and } \mathrm{RCA}_{c,s,t-1} \leq 1\}|, \\
  \mathrm{Exits}_{c,t} &= |\{s : \mathrm{RCA}_{c,s,t} \leq 1 \text{ and } \mathrm{RCA}_{c,s,t-1} > 1\}|, \\
  \mathrm{Net}_{c,t} &= \mathrm{Entries}_{c,t} - \mathrm{Exits}_{c,t}.
\end{align*}

\textbf{Cosine distance} (intensive margin). For each county-year, we represent the skill demand profile as a frequency vector $\mathbf{x}_{c,t} \in \mathbb{R}^{29256}_{+}$ over the full skill vocabulary, then compute
\begin{equation}
  d_{c,t} \;=\; 1 - \frac{\mathbf{x}_{c,t} \cdot \mathbf{x}_{c,t-1}}{\|\mathbf{x}_{c,t}\| \, \|\mathbf{x}_{c,t-1}\|} \in [0, 1].
\end{equation}
Cosine distance is a model-free measure of how much the county's demand profile shifted year-over-year and is computed on raw counts in the 29{,}256-dimensional skill-name space.

\subsection{Entity-type decomposition}
\label{sec:methods:entity}

Our core measures pool postings across all employer entity types in the county-year. The released panel additionally reports decomposed measures along two axes: employer entity type (corporate, university, federal lab, and government), with group I giving each entity's specialization breadth, group J each entity's year-over-year dynamics, and group K pairwise skill similarity between entity-type pairs.

\subsubsection{Entity-type-specific RCA}

For each employer entity type $e \in \{\text{corp}, \text{univ}, \text{fede}, \text{gove}\}$, year $t$, county $c$, and skill $s$, we compute an entity-type-specific RCA from the Phase A \texttt{employer\_skill} table:
\begin{equation}
  \mathrm{RCA}^{e}_{c,s,t} \;=\; \frac{X^{e}_{c,s,t} \,/\, \sum_{s'} X^{e}_{c,s',t}}{\sum_{c'} X^{e}_{c',s,t} \,/\, \sum_{c'}\sum_{s'} X^{e}_{c',s',t}},
\end{equation}
where $X^{e}_{c,s,t}$ counts mentions of skill $s$ in county $c$, year $t$, restricted to postings classified into entity type $e$. The denominator uses the national entity-specific skill pool. Cells with $\sum_{s'} X^{e}_{c,s',t} = 0$ receive $\mathrm{RCA}^{e}_{c,s,t} = 0$. The corresponding entity-type indicator is $M^{e}_{c,s,t} = \mathbbm{1}[\mathrm{RCA}^{e}_{c,s,t} > 1]$. Group I reports $|S^{e}_{c,t}| = \sum_s M^{e}_{c,s,t}$, the entity-type-specific $\mathrm{RCA}>1$ breadth.

\subsubsection{Per-employer-type dynamics (group J)}

For each of the four entity types and each county-year $(c, t)$, we replicate the aggregate group-H churning and cosine-distance measures within the entity type's own skill pool. The entity-type churning measures count skills crossing the $\mathrm{RCA}^{e} = 1$ threshold between consecutive years:
\begin{align*}
  \mathrm{Entries}^{e}_{c,t} &= |\{s : \mathrm{RCA}^{e}_{c,s,t} > 1 \text{ and } \mathrm{RCA}^{e}_{c,s,t-1} \leq 1\}|, \\
  \mathrm{Exits}^{e}_{c,t} &= |\{s : \mathrm{RCA}^{e}_{c,s,t} \leq 1 \text{ and } \mathrm{RCA}^{e}_{c,s,t-1} > 1\}|, \\
  \mathrm{Net}^{e}_{c,t} &= \mathrm{Entries}^{e}_{c,t} - \mathrm{Exits}^{e}_{c,t}.
\end{align*}
The entity-type cosine distance compares the entity-type-specific skill-frequency vectors at consecutive years:
\begin{equation}
  d^{e}_{c,t} \;=\; 1 - \frac{\mathbf{x}^{e}_{c,t} \cdot \mathbf{x}^{e}_{c,t-1}}{\|\mathbf{x}^{e}_{c,t}\| \, \|\mathbf{x}^{e}_{c,t-1}\|} \in [0, 1].
\end{equation}
Group-J measures are noisier than the aggregate group-H counterparts in county-years with low entity-type posting volume. The codebook documents the recommended posting-count thresholds.

\subsubsection{Employer-pair skill similarity (group K)}

For each pair of entity types $(a, b) \in \{(\text{univ}, \text{corp}), (\text{fede}, \text{corp}), (\text{gove}, \text{corp}), (\text{univ}, \text{fede}), (\text{univ}, \text{gove}), (\text{fede}, \text{gove})\}$ and each county-year $(c, t)$, we compute six families of similarity measures. Each family is computed pooled across all skills and separately over each skill-type subset $\tau \in \{\text{all}, \text{specialized}, \text{software}, \text{common}\}$, producing $6 \text{ pairs} \times 6 \text{ families} \times 4 \text{ skill-type splits} = 144$ variables in group K.

\textbf{Cosine similarity.}\quad The alignment of the two entity-type-specific skill-frequency vectors $\mathbf{x}^{a,\tau}_{c,t}$ and $\mathbf{x}^{b,\tau}_{c,t}$:
\begin{equation}
  \mathrm{cosine}^{\tau}_{a,b,c,t} \;=\; \frac{\mathbf{x}^{a,\tau}_{c,t} \cdot \mathbf{x}^{b,\tau}_{c,t}}{\|\mathbf{x}^{a,\tau}_{c,t}\| \, \|\mathbf{x}^{b,\tau}_{c,t}\|} \in [0,1].
\end{equation}

\textbf{Jaccard.}\quad Overlap between the entity-type-specific $\mathrm{RCA}>1$ skill sets $S^{a,\tau}_{c,t}$ and $S^{b,\tau}_{c,t}$:
\begin{equation}
  \mathrm{jaccard}^{\tau}_{a,b,c,t} \;=\; \frac{|S^{a,\tau}_{c,t} \cap S^{b,\tau}_{c,t}|}{|S^{a,\tau}_{c,t} \cup S^{b,\tau}_{c,t}|}.
\end{equation}

\textbf{Hidalgo technological proximity.}\quad Average pairwise skill-space relatedness between the two entity-type RCA portfolios \citep{hidalgo2007product}, using the year-specific co-occurrence matrix $\Phi_t$ from \Cref{sec:methods:phaseb}:
\begin{equation}
  \mathrm{hidalgo}^{\tau}_{a,b,c,t} \;=\; \frac{1}{|S^{a,\tau}_{c,t}| \, |S^{b,\tau}_{c,t}|} \sum_{s \in S^{a,\tau}_{c,t}} \sum_{s' \in S^{b,\tau}_{c,t}} \phi^{\text{co}}_{s,s',t}.
\end{equation}
This captures nearby skills in skill-space, not only exact overlap.

\textbf{Weighted RCA overlap.}\quad Average $b$-sector RCA evaluated across the $a$-sector specialization set:
\begin{equation}
  \mathrm{rca\_overlap}^{\tau}_{a,b,c,t} \;=\; \frac{1}{|S^{a,\tau}_{c,t}|} \sum_{s \in S^{a,\tau}_{c,t}} \mathrm{RCA}^{b,\tau}_{c,s,t}.
\end{equation}
High values mean the $a$ sector demands skills the $b$ sector already specializes in.

\textbf{Directional skill gap, count and relatedness.}\quad For each $(a,b)$ pair, define the gap skill set $G^{a,\tau}_{b,c,t} = S^{a,\tau}_{c,t} \setminus S^{b,\tau}_{c,t}$: $a$-sector specializations that the $b$ sector lacks $\mathrm{RCA}>1$ on. The gap count is $|G^{a,\tau}_{b,c,t}|$. The gap relatedness measures how close the gap skills are to the $b$ sector's current portfolio:
\begin{equation}
  \mathrm{gap\_rel}^{\tau}_{a,b,c,t} \;=\; \frac{1}{|G^{a,\tau}_{b,c,t}| \, |S^{b,\tau}_{c,t}|} \sum_{s \in G^{a,\tau}_{b,c,t}} \sum_{s' \in S^{b,\tau}_{c,t}} \phi^{\text{co}}_{s,s',t}.
\end{equation}
The two gap measures are directional: $\mathrm{gap}^{\tau}_{a,b}$ in general differs from $\mathrm{gap}^{\tau}_{b,a}$.

\subsection{Compute environment}

The full pipeline runs on a single compute node of the ASU Sol HPC cluster (Arizona State University's research supercomputer) and completes in approximately 47 hours of wall time. Phase A's full-data scan dominates the runtime; Phase B's measure computation finishes in about 30 minutes from the per-year intermediates. SLURM job scripts and exact resource allocations are included in the public code repository (\Cref{sec:code}). The pipeline is deterministic given the same input files and code version.

\section{Data Records}
\label{sec:records}

The released dataset is the county-year panel of skill specialization, relatedness, and complexity, across employer entity type. It is distributed as a single Apache Parquet file, \texttt{county\_year\_panel.parquet}, with 47{,}891 county-year observations and 201 variables. The variables are organized into eleven groups (A--K). The 37-variable core set (groups A--H) reports aggregate county-year measures sufficient for most descriptive, teaching, and applied uses. The 164-variable entity-decomposed extension adds entity-specific measures of each entity type's own specialization and dynamics (groups I and J), plus entity-pair measures of alignment, overlap, and skill gaps between entity types within the same county-year (group K). Appendix Table~\ref{tab:variable_groups} outlines and defines every group; Appendix Table~\ref{tab:summary_stats} reports variable-level summary statistics for the 35 substantive core variables in groups B--H (excluding the two identifiers in group A). A CSV mirror, \texttt{county\_year\_panel.csv}, is provided for users without parquet support.

A data dictionary, codebook, and reproducibility README accompany the 
data in the repository. The codebook documents every variable in detail, 
flags the core set, and shows how to load only that subset. Parquet is 
the recommended format for analytic workflows in pandas, polars, R 
(\texttt{arrow}), Stata 18+, Julia (\texttt{Parquet.jl}), and DuckDB. The 
parquet file is compressed with zstd, level 22.

\subsection{Primary record: county-year panel}

The core set documents each of the three conceptual dimensions
introduced in \Cref{sec:background}.

\textbf{(A) Unit identifiers (2 variables).}\quad 5-digit county FIPS and calendar year.

\textbf{(B) Labor demand: posting and skill-mention totals (5 variables).}\quad Total postings, postings with at least one parsed skill, and the total mention counts of specialized, software, and common skills.

\textbf{(C) Labor demand: entity-type posting counts (4 variables).}\quad Posting counts in each of the four entity types: corporate (all private-sector postings), university, federal lab, and government.

\textbf{(D) Nature of work: modality and internships (7 variables).}\quad Posting counts and shares for remote, hybrid, and on-site postings; the count of internship-flagged postings.

\textbf{(E) Skill composition (5 variables).}\quad The three skill-type shares (specialized, software, and common), mean skill mentions per posting, and the percent of postings with at least one parsed skill.

\textbf{(F) Skill diversity, concentration, and complexity (7 variables).}\quad Count of distinct skills, count of $\mathrm{RCA}>1$ skills, average ubiquity of the $\mathrm{RCA}>1$ portfolio, Herfindahl-Hirschman concentration (HHI), Shannon entropy, ECI, and Tacchella fitness-complexity.

\textbf{(G) Skill relatedness and network position (3 variables).}\quad Balland skill density, Neffke skill coherence, and the average network centrality of the county's $\mathrm{RCA}>1$ skills in the skill-space network.

\textbf{(H) Year-over-year dynamics (4 variables).}\quad Counts of skills that gained and lost $\mathrm{RCA}>1$ relative to the prior year (entries, exits, and net), and cosine distance between the current and prior-year skill frequency vectors.

\subsection{Entity-decomposed extension: groups I, J, and K}
\label{sec:records:extended}

The remaining 164 variables decompose the entity-type aspect of skill demand along two axes. The entity-specific measures (groups I and J) characterize each entity type's own specialization and dynamics, in parallel to the aggregate measures in groups F--H. The entity-pair measures (group K) characterize the relationship between any two entity types within the same county-year: alignment of their skill demands, overlap of their specializations, and the directional skill gaps between them. Full construction details and equations for each measure appear in \Cref{sec:methods:entity}; we summarize the operational definitions here.

\textbf{Entity-type-specific RCA.}\quad The entity-specific and entity-pair measures all build on an entity-type-specific Balassa RCA. For each entity type $e \in \{\text{corp}, \text{univ}, \text{fede}, \text{gove}\}$, county $c$, year $t$, and skill $s$, $\mathrm{RCA}^{e}_{c,s,t}$ is the share of skill $s$ in entity $e$'s skill mentions in county $c$, divided by the share of skill $s$ in entity $e$'s national skill mentions, both in year $t$. The binary indicator $M^{e}_{c,s,t} = \mathbbm{1}[\mathrm{RCA}^{e}_{c,s,t} > 1]$ marks each skill where entity $e$ over-specializes in county $c$ relative to entity $e$'s national pool. This restricts the comparison set so that, for example, university specializations are benchmarked against the national university skill pool rather than the all-employers pool, isolating the sector-specific dimension of specialization.

\textbf{(I) Entity-type specialization breadth (4 variables; entity-specific).}\quad For each entity type $e$ and county-year, the count $\sum_s M^{e}_{c,s,t}$ of skills with $\mathrm{RCA}^{e} > 1$. The four released variables are \texttt{corp\_n\_rca\_skills}, \texttt{univ\_n\_rca\_skills}, \texttt{fede\_n\_rca\_skills}, and \texttt{gove\_n\_rca\_skills}. Group I asks: how broadly does each entity type specialize within each county-year?

\textbf{(J) Per-employer-type skill dynamics (16 variables; entity-specific).}\quad For each entity type $e$, we compute four year-over-year measures in parallel to the aggregate group-H measures, but on $e$'s own skill pool. \textbf{Entries} counts skills that crossed into $\mathrm{RCA}^{e} > 1$ this year relative to $t-1$; \textbf{exits} counts skills that fall below $\mathrm{RCA}^{e} = 1$; \textbf{net} is entries minus exits. \textbf{Cosine distance} is $1 - \cos(\mathbf{x}^{e}_{c,t}, \mathbf{x}^{e}_{c,t-1})$ on entity $e$'s 29{,}256-dimensional skill-frequency vectors. Group J separates sector-specific shifts from aggregate shifts: when group H reports churn in a county-year, group J reveals which entity type drove it. These measures are noisier than the aggregate group-H counterparts in county-years with low entity-type posting volume.

\textbf{(K) Employer-pair skill similarity (144 variables; entity-pair).}\quad For each of the six unordered entity-type pairs $(a,b)$ (univ-corp, fede-corp, gove-corp, univ-fede, univ-gove, and fede-gove), we compute six measure families. The first three operationalize standard concepts from the economic-geography and similarity literatures. \textbf{Cosine similarity} between the entity-$a$ and entity-$b$ skill-frequency vectors measures how aligned their demand profiles are. \textbf{Jaccard} on the entity-specific $\mathrm{RCA}>1$ skill sets ($|S^a \cap S^b| / |S^a \cup S^b|$) measures the exact overlap of their specializations. \textbf{Hidalgo} \cite{hidalgo2007product} technological proximity is the average pairwise skill-skill relatedness $\phi^{\text{co}}_{s,s'}$ between $S^a$ and $S^b$ using the national skill-space relatedness matrix from \Cref{sec:methods:phaseb}; this captures nearby skills, not only exact overlap. The remaining three measures are constructed for this dataset to capture asymmetric specialization relationships. \textbf{Weighted RCA overlap} reports the average $\mathrm{RCA}^{b}$ across the $a$-specialization set $S^a$, indicating how strongly entity $b$ specializes in the skills entity $a$ specializes in. The two directional gap measures restrict to skills $a$ specializes in but $b$ does not. \textbf{Gap count} is $|S^a \setminus S^b|$. \textbf{Gap relatedness} is the average $\phi^{\text{co}}$ between the gap skills and entity $b$'s own RCA portfolio, indicating how close $b$ is to acquiring those gap skills. Each measure is computed in four versions: pooled across all skills and separately over specialized, software, and common skill subsets, yielding $6 \text{ pairs} \times 6 \text{ families} \times 4 \text{ skill-type splits} = 144$ variables.



\subsection{File format and access}

The parquet file uses zstd level 22 compression for maximum compression at acceptable read speed. The parquet is approximately 30 MB on disk; the CSV mirror is approximately 85 MB.

For users who prefer non-parquet formats, the \texttt{pyarrow} and \texttt{polars} Python packages, the \texttt{arrow} R package, and DuckDB all read parquet directly. Stata 18 and later read parquet via \texttt{frame import}. The CSV mirror is provided for users without parquet support.

\subsection{Data hosting and license}

The county-year panel and its accompanying README, data dictionary, and other information are deposited at \url{https://doi.org/10.5281/zenodo.20330086} and additionally mirrored at the project GitHub repository \url{https://github.com/AntJam-Howell/skills-econ-geog-data}.  See the Code Availability statement at the end of the paper for source code locations. The released panel is distributed under the Creative Commons Attribution 4.0 International License (CC~BY~4.0). The full license text accompanies the data in both the DOI deposit and the GitHub mirror; downstream users may copy, redistribute, and adapt the data, including for commercial purposes, provided they cite this article and the deposit DOI.

\subsection{Interactive dashboard}
\label{sec:records:dashboard}

This article accompanies an interactive web dashboard that visualizes the released county-year panel. The dashboard is hosted at \url{https://skills-econ-geog.netlify.app/} and is intended for readers, students, and policy users who want to explore the data without writing code. It is open access and does not require credentials.  The dashboard reads the released county-year panel directly and focuses on the core economic-geography measures: specialization, concentration, diversity, complexity, relatedness, and overall posting volume. This scope is designed to support descriptive exploration, teaching, and applied use without requiring readers to navigate the full 201-variable schema.

The dashboard is organized into five pages: (i) \textbf{Spatial visualization}, a county-level choropleth of any core measure with a year slider and play button to animate 2010--2024; (ii) \textbf{Rankings \& trends}, the top-25 ranked counties for a selected measure and year, a distribution histogram, and four national-context time-series charts; (iii) \textbf{County comparisons}, bivariate scatter exploration of any two core measures for a selected year, with a focal county and its k-nearest peers highlighted; (iv) \textbf{County profiles}, an in-depth single-county trajectory across the full 15-year window, with sparklines and stacked composition plots for the work-mode and skill-type shares; and (v) \textbf{How to use the dashboard}, documentation of the core measures, end-to-end workflows, and methodology notes.

\section{Technical Validation}
\label{sec:validation}

This section validates the released data along three axes: (i) external benchmarking against published vacancy statistics and against expected geographic, employer-type, and spatial gradients; (ii) replication of expected theoretical properties of the diversity measures; and (iii) bivariate consistency between key measures.

\subsection{Coverage and external benchmarking}

\begin{figure}[H]
\centering
\includegraphics[width=0.95\textwidth]{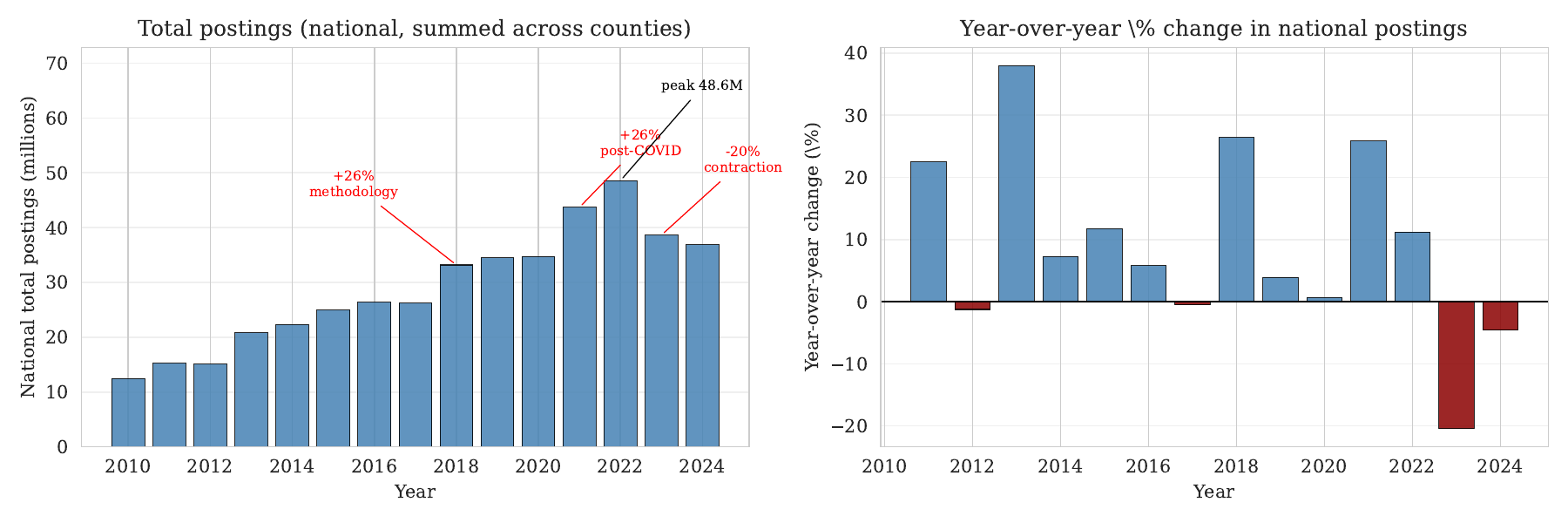}
\caption{National posting volumes, summed across counties (left), and year-over-year percent change (right). The visible features (2017--2018 coverage step-up, no 2020 dip, post-2022 contraction) are validated against external sources: Lightcast release notes for the 2018 step-up, and BLS JOLTS for the COVID-period pattern.}
\label{fig:national_totals}
\end{figure}

\Cref{fig:national_totals} shows national posting volumes summed across all counties, with year-over-year percent changes in the right panel. Three structural features are visible and warrant comment.

\textbf{2017--2018 step-up.}\quad National postings increase by 26 percent between 2017 and 2018, from 26 million to 33 million. This jump reflects a Lightcast coverage-methodology change documented in the vendor's release notes: new sources were added to the crawl and the parser was upgraded. The level shift is not an artifact of the present pipeline; it is present in the raw vendor data. Analysts should use posting shares rather than levels when comparing across the 2017--2018 boundary.

\textbf{No COVID dip.}\quad Posting activity was essentially flat from 2019 to 2020 ($+0.7$ percent at the national aggregate). This is consistent with evidence from the BLS Job Openings and Labor Turnover Survey (JOLTS) that vacancy postings remained elevated through 2020 even when hiring slowed; firms maintained job advertisements through the pandemic while extending time-to-fill \citep{bls2024jolts}.

\textbf{Post-COVID surge then contraction.}\quad Postings jumped 26 percent in 2021, peaked at 48.6 million in 2022, contracted 20 percent in 2023, and contracted a further 5 percent in 2024. This is consistent with the JOLTS time series and with the contemporaneous policy discussion of the post-pandemic labor-market cooling.

\begin{figure}[H]
\centering
\includegraphics[width=0.95\textwidth]{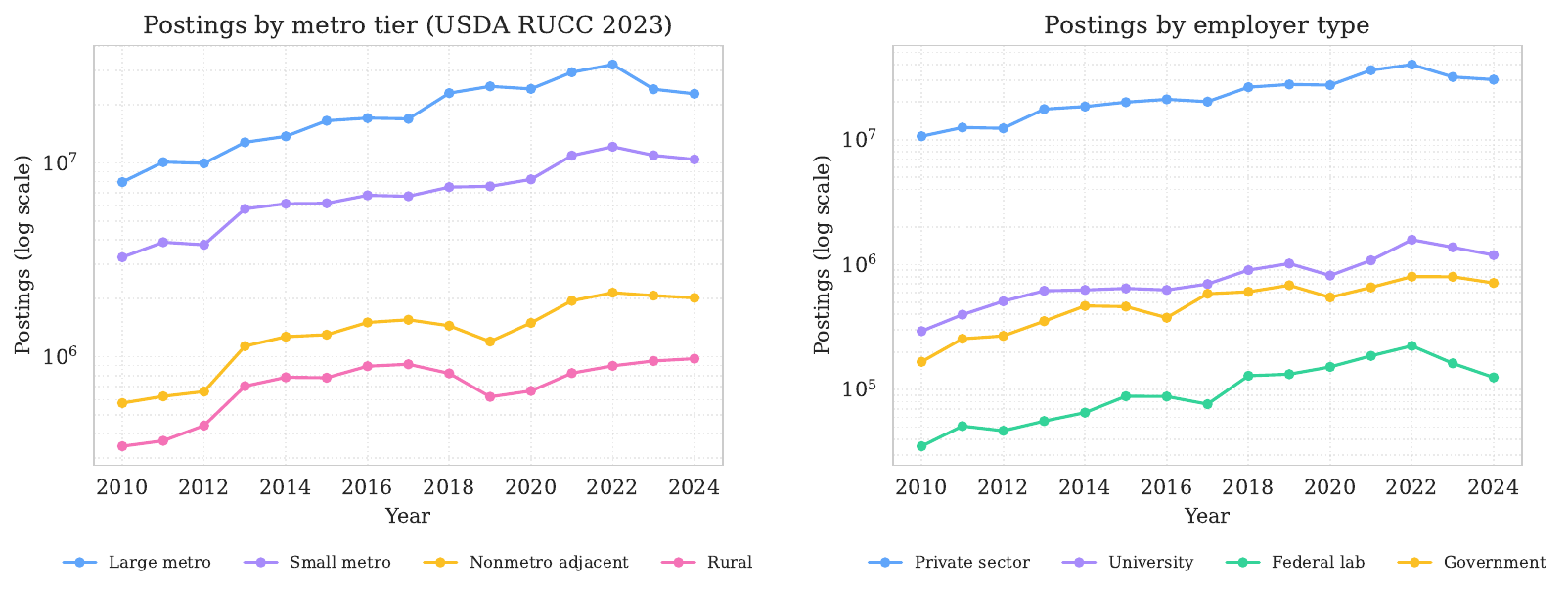}
\caption{National posting volume by metro tier (left) and employer type (right), log scale.}
\label{fig:coverage_geo_employer}
\end{figure}

\Cref{fig:coverage_geo_employer} disaggregates national posting volumes along two cuts. The left panel shows postings by USDA Rural-Urban Continuum Code (RUCC 2023) metro tier, using the standard 4-tier collapse: large metro (RUCC 1), small metro (RUCC 2--3), nonmetro adjacent (RUCC 4, 6, 8), and rural (RUCC 5, 7, 9). State-level FIPS codes ending in 999 lack a county-level RUCC assignment and are excluded from this panel. The right panel shows postings by employer entity type: corporate company in the private sector, university, federal lab, and government.

Two qualitative patterns serve as external validation. Posting volume falls monotonically across the metro-tier gradient. Large-metro counties carry roughly an order of magnitude more postings than small metros, and another order of magnitude more than rural counties. This matches the well-established geographic concentration of U.S.~employment in larger metropolitan economies. The four employer types also separate cleanly by scale: private-sector postings sit two orders of magnitude above university and government postings, and roughly three orders of magnitude above federal labs.

\begin{figure}[H]
\centering
\includegraphics[width=0.7\textwidth]{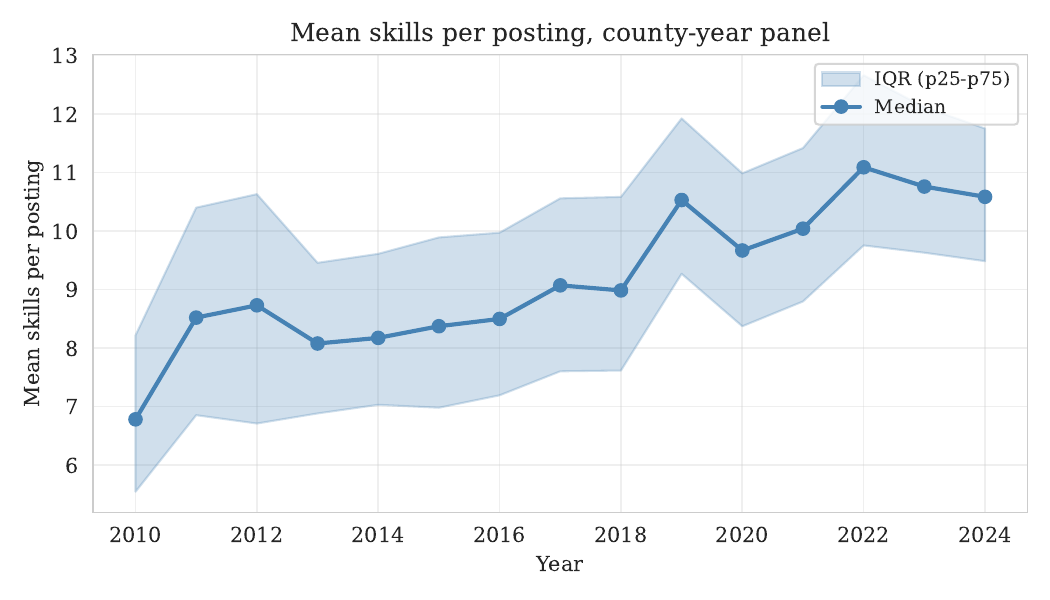}
\caption{Mean skill mentions per posting at the median county, with interquartile range. The trend reflects both an expanding skill taxonomy and longer stated skill lists per posting.}
\label{fig:mean_skills}
\end{figure}

\Cref{fig:mean_skills} shows the mean number of skill mentions per posting at the median county. This rises from approximately 6.8 in 2010 to a peak of 11.1 in 2022, then eases modestly to roughly 10.6 by 2024. The national aggregate is somewhat higher (11 rising to 14) because higher-volume counties also post more skills per posting on average. Both series indicate a broadening of stated skill requirements over time, consistent with the documented expansion of the Lightcast skill taxonomy and with the trend toward longer, more-skill-listed postings in the underlying data \citep{deming2018skill}.

\begin{figure}[H]
\centering
\includegraphics[width=0.85\textwidth]{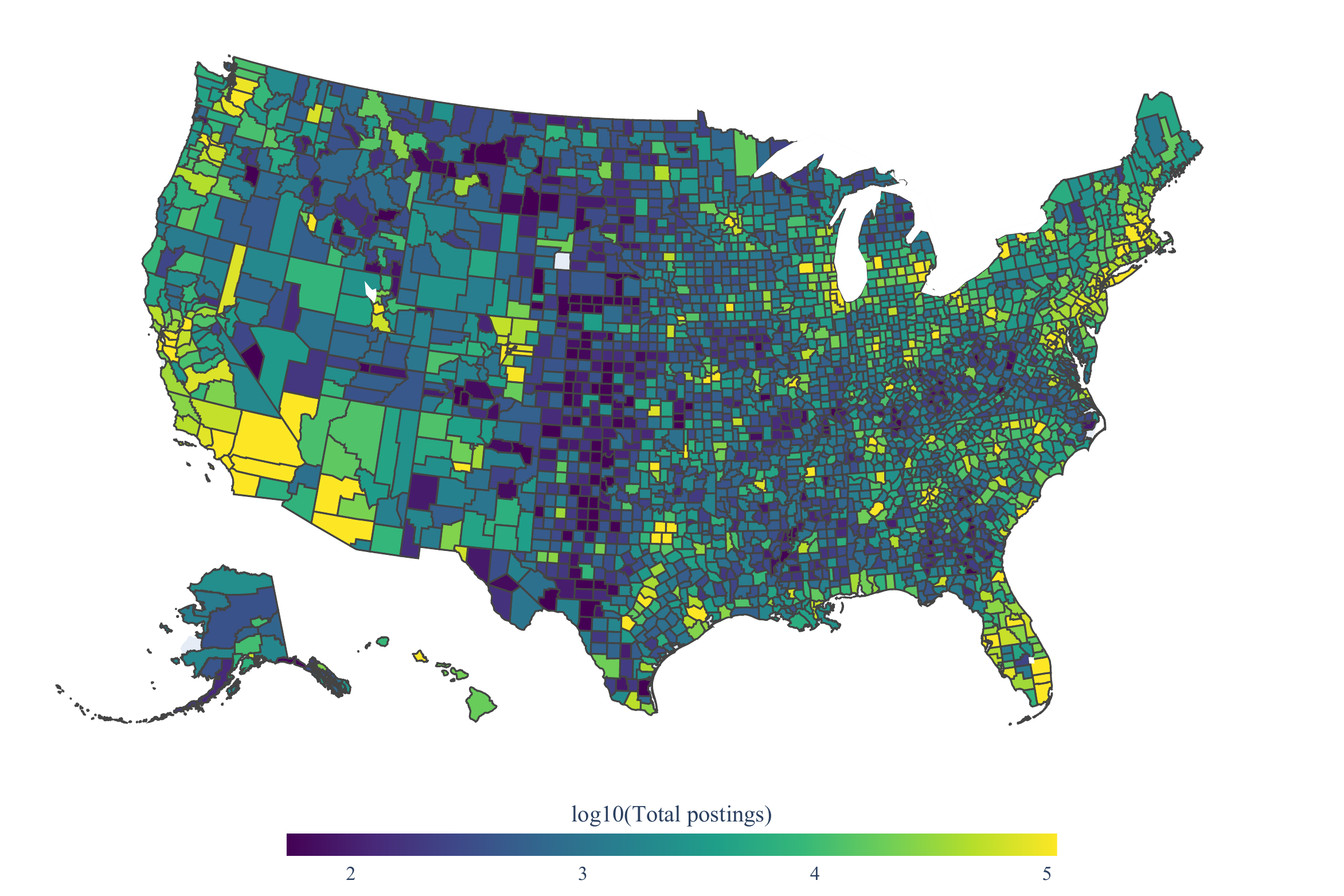}
\caption{Total postings by county, 2024 (log color scale).}
\label{fig:map_total_postings}
\end{figure}

\begin{figure}[H]
\centering
\includegraphics[width=0.85\textwidth]{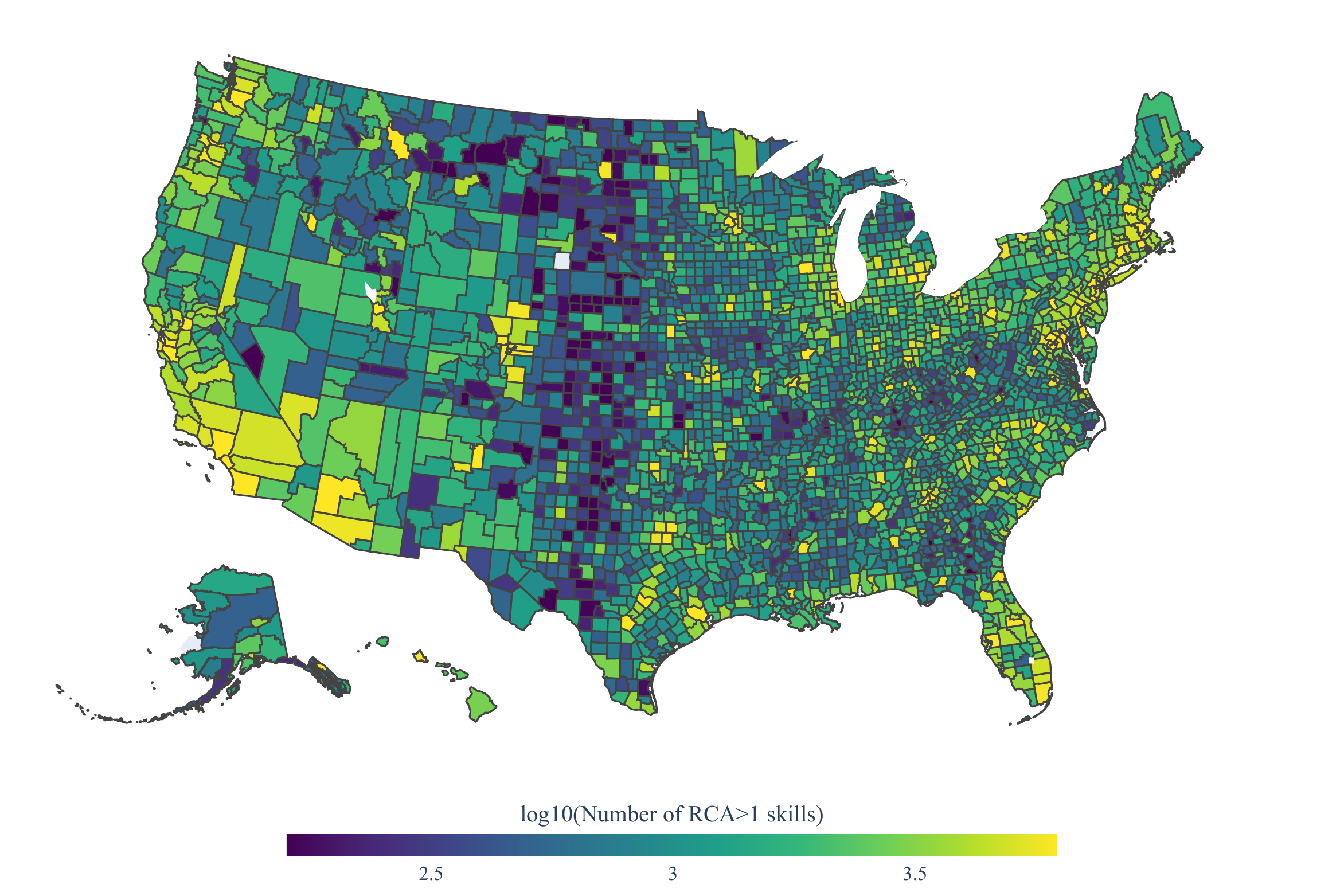}
\caption{Number of $\mathrm{RCA}>1$ skills by county, 2024 (log color scale).}
\label{fig:map_n_rca_skills}
\end{figure}

\begin{figure}[H]
\centering
\includegraphics[width=0.85\textwidth]{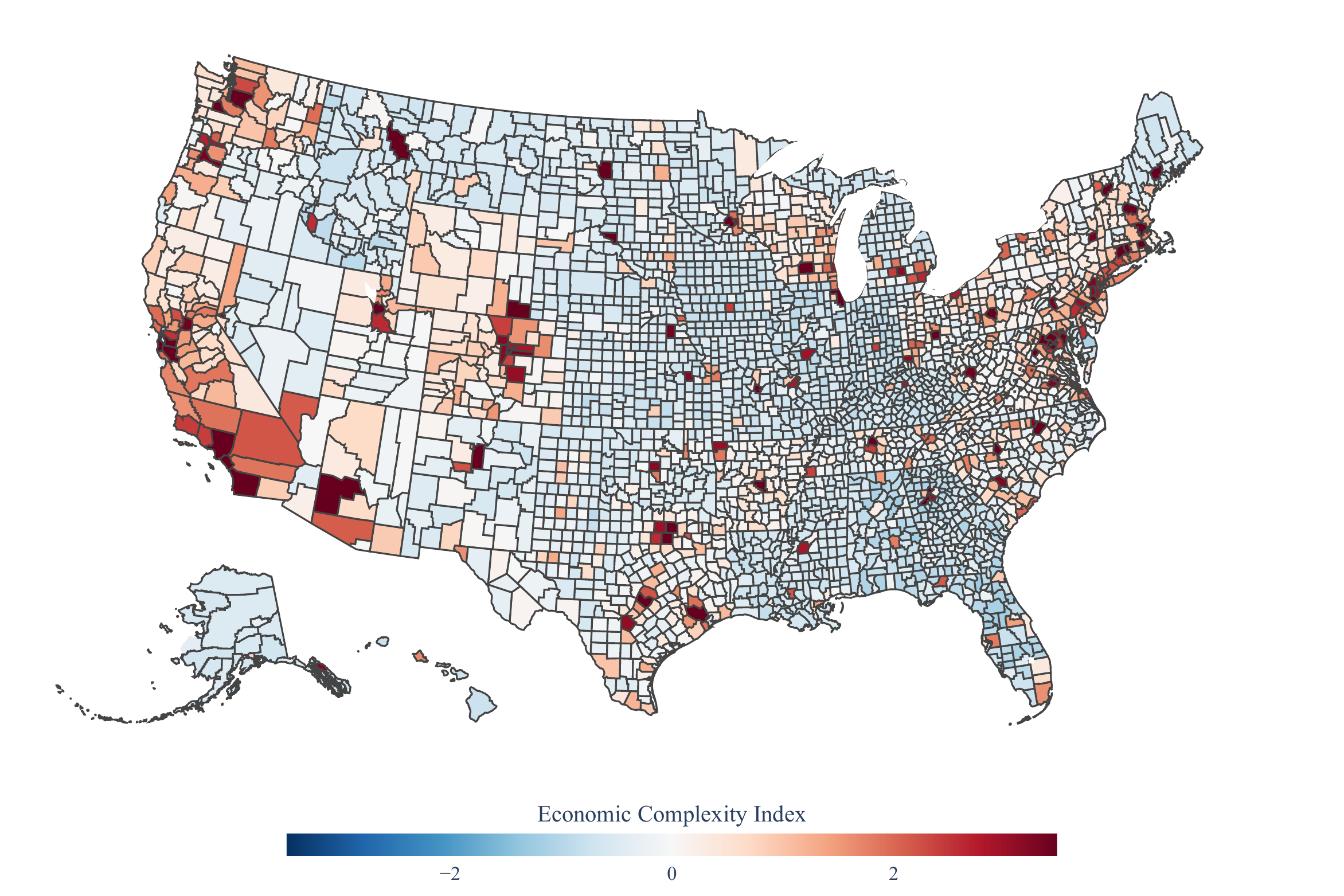}
\caption{ECI by county, 2024 (diverging color scale).}
\label{fig:map_eci}
\end{figure}

The cross-sectional geography of skill demand provides a further coverage check. \Cref{fig:map_total_postings,fig:map_n_rca_skills,fig:map_eci} map total postings, the count of $\mathrm{RCA}>1$ skills, and the ECI by county in 2024. Posting volume concentrates in the well-known U.S.~labor-market hubs (the Bay Area, Los Angeles, Seattle, the Boston--New York--Washington corridor, Chicago, Atlanta, Dallas, Houston, Phoenix, and the Florida metros). The count of $\mathrm{RCA}>1$ skills traces the same urban-research gradient, with the largest metros showing the broadest specialization portfolios. The ECI, on a diverging scale, places coastal metros, the Boston--Washington corridor, and major university towns in the upper tail and rural counties in the South, Midwest, and Mountain West in the lower tail. 

\subsection{Testing Economic Geography Predictions}

\begin{figure}[H]
\centering
\begin{minipage}{0.48\textwidth}
\centering
\includegraphics[width=\textwidth]{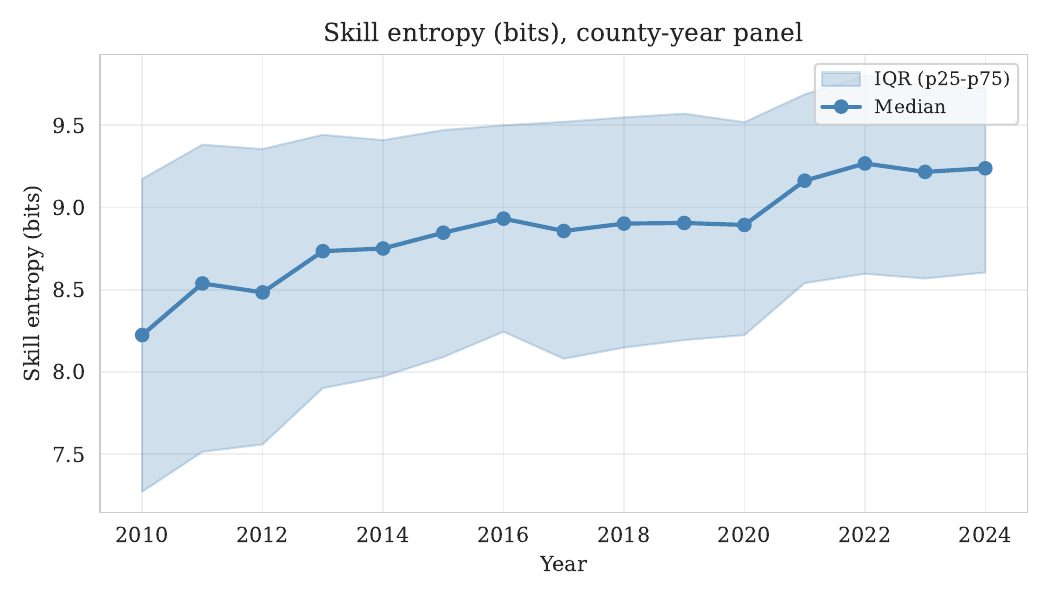}
\caption{Shannon entropy of skill mentions, median county with IQR.}
\label{fig:entropy}
\end{minipage}%
\hfill
\begin{minipage}{0.48\textwidth}
\centering
\includegraphics[width=\textwidth]{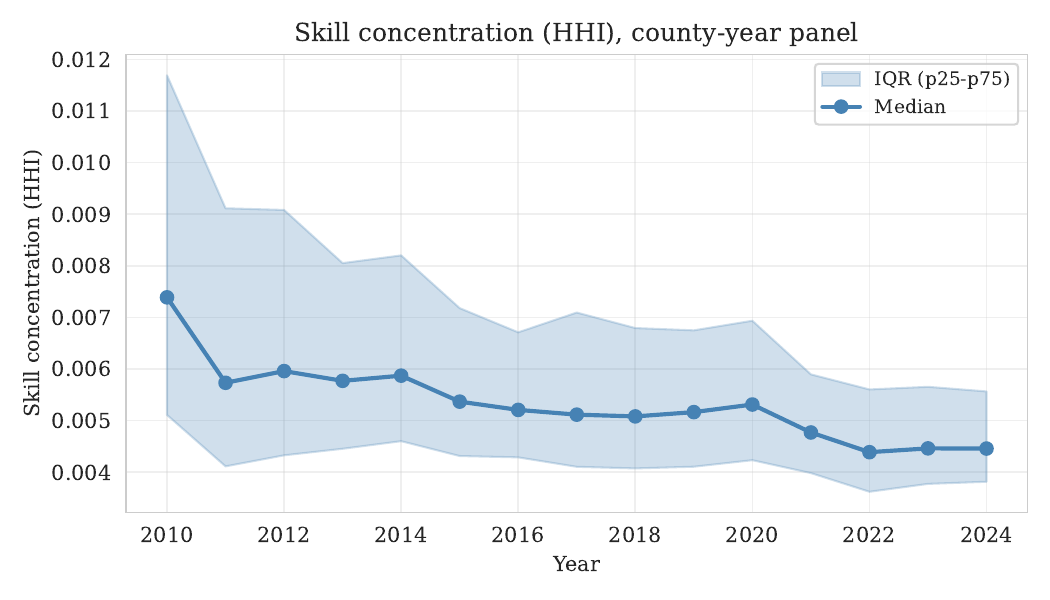}
\caption{HHI of skill mentions, median county with IQR.}
\label{fig:hhi}
\end{minipage}
\end{figure}

The released measures are constructed from different transformations of the underlying county-skill-year matrix, but their cross-sectional relationships should reproduce well-known empirical regularities of local labor markets.  First, the economic geography literature predicts that as the skill taxonomy expands and counties diversify their posted skill requirements, Shannon entropy of skill mentions should rise and HHI should fall. \Cref{fig:entropy,fig:hhi} confirm this pattern for the median county, with entropy trending up and HHI trending down across the 15-year window.

\begin{figure}[H]
\centering
\includegraphics[width=0.7\textwidth]{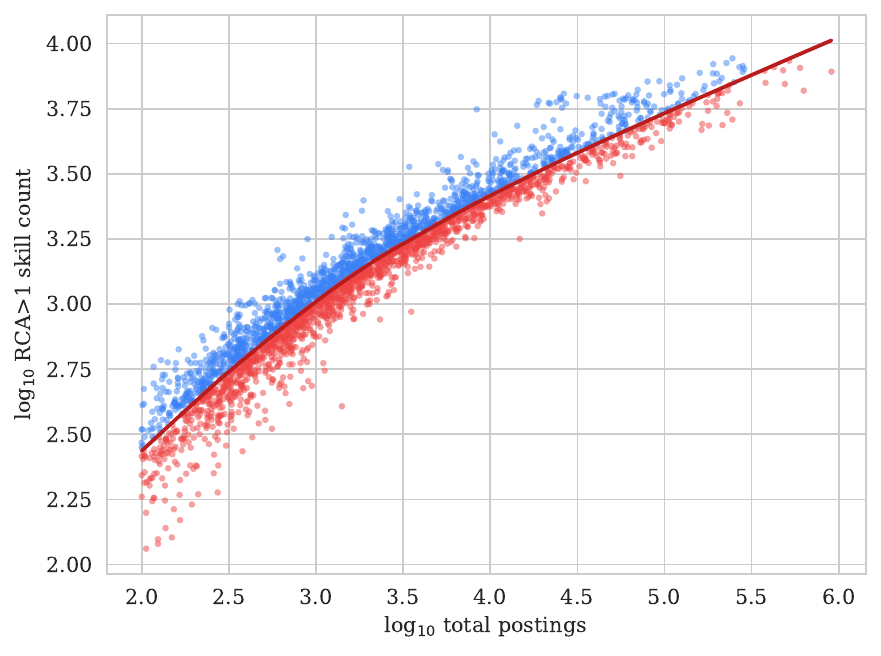}
\caption{County count of $\mathrm{RCA}>1$ skills versus total postings, 2024 (log-log). \\ \textit{Notes:} Observations are counties. The sample is the $n = 2{,}957$ counties with at least 100 postings in 2024, excluding state-level placeholder FIPS codes ending in 999. The fitted curve is a LOWESS smoother (bandwidth 0.3) of $\log_{10}$ RCA-skill count on $\log_{10}$ total postings; counties are colored blue if above the curve, red if below. Spearman rank correlation between $\log_{10}$ total postings and $\log_{10}$ RCA-skill count is $\rho = 0.976$. The Spearman statistic is reported in preference to Pearson because the fitted curve is a LOWESS smoother of a nonlinear monotonic relationship.}
\label{fig:scatter_postings_rca}
\end{figure}

Next, we inspect three bivariate scatters on U.S.~counties with at least 100 postings in 2024, excluding the 51 state-level placeholder FIPS codes that end in 999. The observations in each scatter are counties. Each scatter plots one panel variable against another, fits a regression curve in the displayed-axis space (LOWESS for the postings vs.~$\mathrm{RCA}>1$ and density vs.~coherence pairs; OLS in log-y space for ECI vs.~fitness), and colors counties blue if they sit above the curve and red if they sit below. The first scatter pairs total postings with the count of $\mathrm{RCA}>1$ skills on a log-log scale (\Cref{fig:scatter_postings_rca}): counties with larger labor markets specialize in more skills, as expected from the breadth-mass relationship that motivates the economic-complexity literature.

\begin{figure}[H]
\centering
\includegraphics[width=0.7\textwidth]{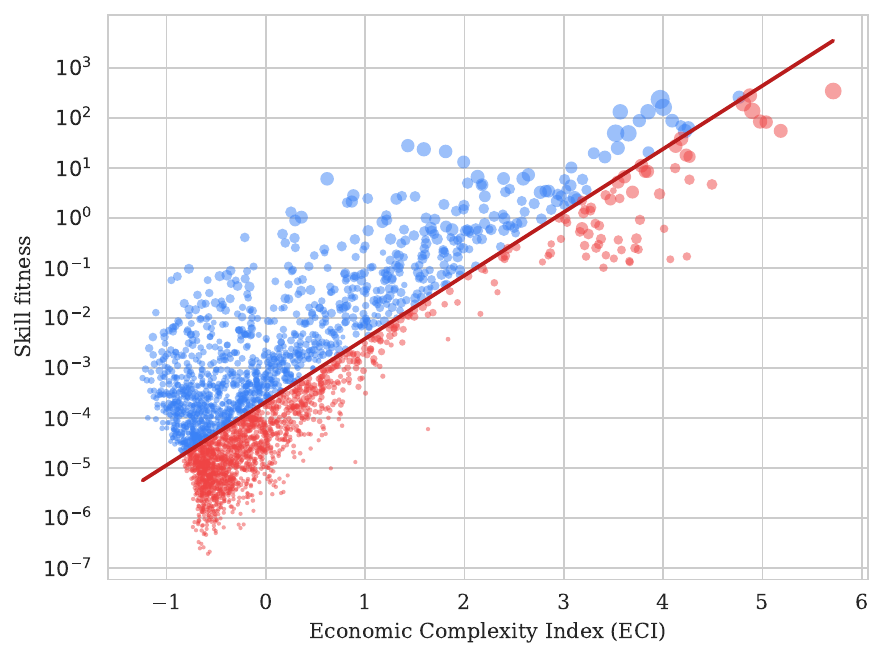}
\caption{Tacchella skill fitness (log scale) versus ECI, 2024. \\ \textit{Notes:} Observations are counties. The sample is the $n = 2{,}957$ counties with at least 100 postings in 2024, excluding state-level placeholder FIPS codes ending in 999. An OLS line is fitted in (ECI, $\log_{10}$ fitness) space; counties are colored blue if above the line, red if below. Pearson correlation in the displayed (ECI, $\log_{10}$ fitness) space is $r = 0.775$.}
\label{fig:scatter_eci_fitness}
\end{figure}

\begin{figure}[H]
\centering
\includegraphics[width=0.7\textwidth]{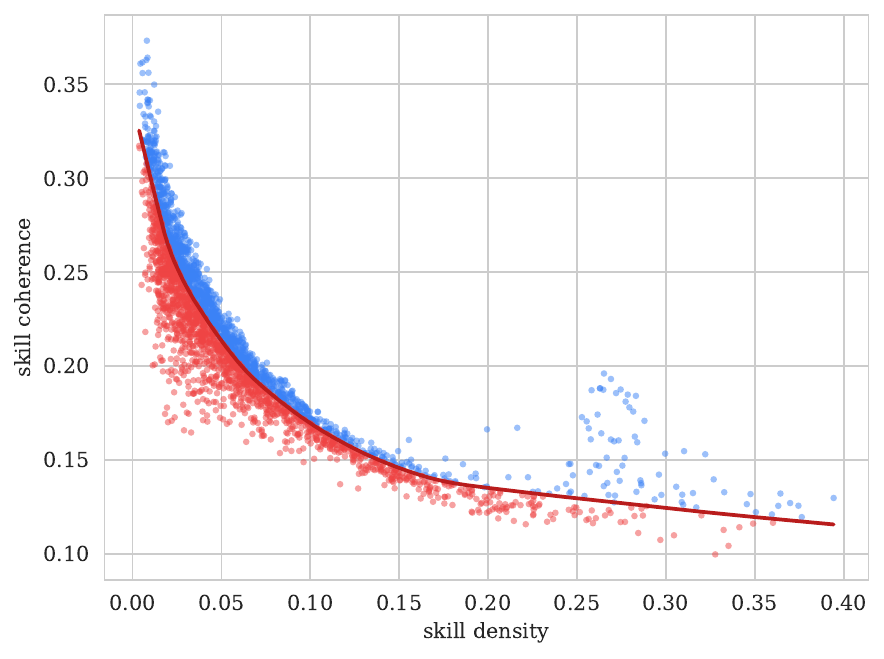}
\caption{Skill coherence versus skill density, 2024. \\ \textit{Notes:} Observations are counties. The sample is the $n = 2{,}957$ counties with at least 100 postings in 2024, excluding state-level placeholder FIPS codes ending in 999. The fitted curve is a LOWESS smoother (bandwidth 0.3); counties are colored blue if above the curve, red if below. Spearman rank correlation is $\rho = -0.932$. The Spearman statistic is reported in preference to Pearson because the fitted curve is a LOWESS smoother of a nonlinear monotonic relationship.}
\label{fig:scatter_coherence_density}
\end{figure}

The second scatter pairs the ECI with fitness-complexity on a log-y axis (\Cref{fig:scatter_eci_fitness}). Both target complexity but with different aggregation logics (linear method-of-reflections versus nonlinear iteration); the two indices rank counties consistently in the bulk of the distribution while leaving room for divergence in the tails. The third scatter pairs skill density with skill coherence (\Cref{fig:scatter_coherence_density}). The two measure complementary aspects of a county's specialization portfolio: density is the average proximity of the county's skills to other potential specializations, while coherence is the internal relatedness of the existing portfolio. The relationship is strongly negative and visibly curved, consistent with a trade-off in which counties with broadly proximal portfolios (high density) tend to have less internally clustered specializations (low coherence), and vice versa.

\subsection{Entity-decomposed extension: stocks, flows, and alignment over time}

\begin{figure}[H]
\centering
\includegraphics[width=0.95\textwidth]{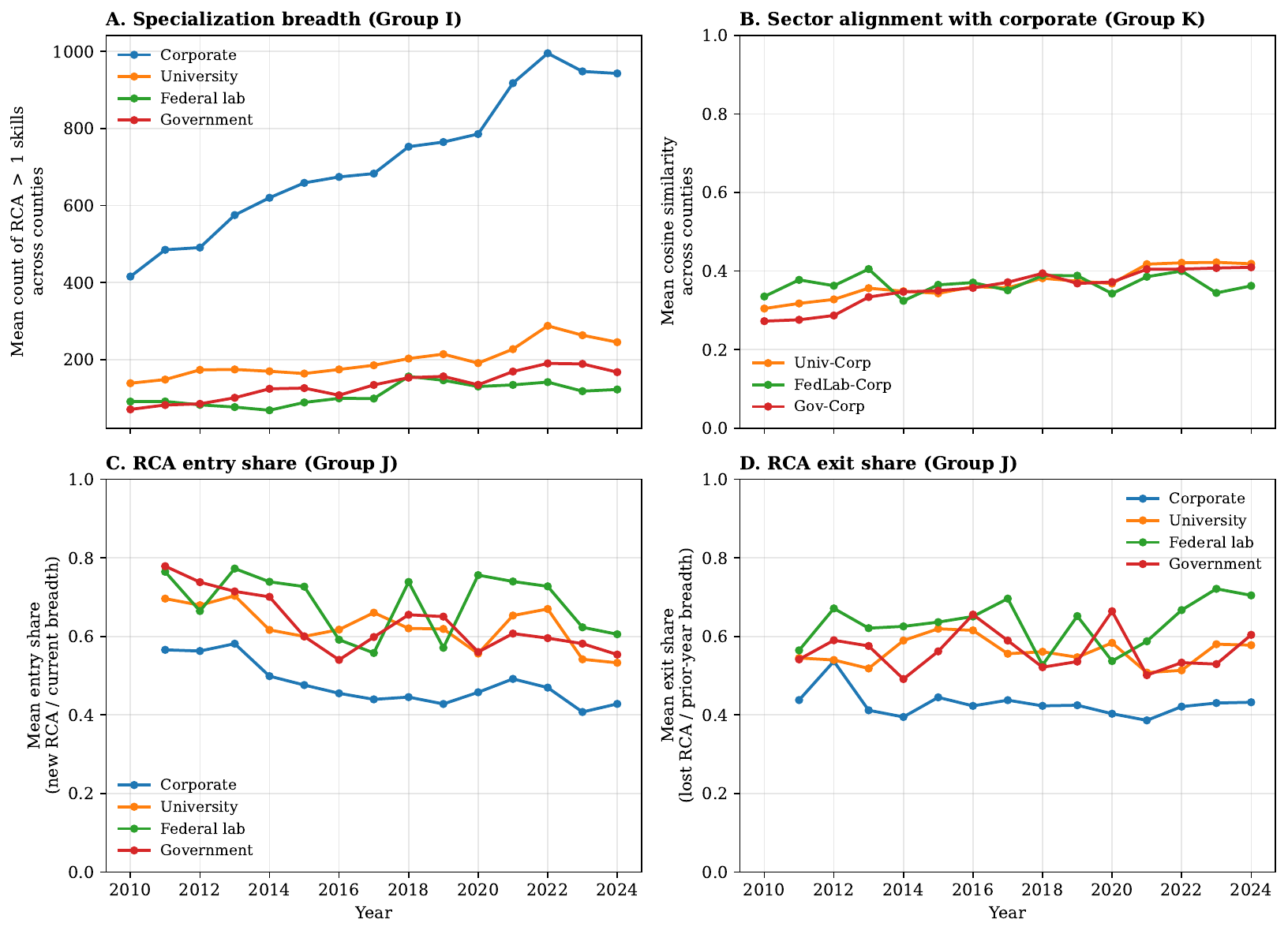}
\caption{Entity-decomposed extension over 2010--2024. \\ \textit{Notes:} \textbf{A}: Group I, mean within-entity count of skills at $\mathrm{RCA}>1$, by entity type. \textbf{B}: Group K, mean cosine similarity between each public-sector entity and corporate. \textbf{C}: Group J, mean RCA entry share, computed as the count of skills newly at $\mathrm{RCA}>1$ in year $t$ divided by the entity's breadth in year $t$. \textbf{D}: Group J, mean RCA exit share, computed as the count of skills that lost $\mathrm{RCA}>1$ in year $t$ divided by the entity's breadth in year $t-1$. Each line shows the national mean computed across counties where the variable is defined for that year (an entity must have at least one posting in the county-year for its measures to be defined; Group K additionally requires both entities of the pair to be present).}
\label{fig:entity_extension_overview}
\end{figure}

The 164 variables of the entity-decomposed extension report parallel measures for each of the four employer entity types. \Cref{fig:entity_extension_overview} demonstrates that these variables behave coherently across the 2010--2024 window along the three axes the extension is designed to span: the \textbf{stock} of specializations within each entity type (Group I), the \textbf{between-entity alignment} of skill demands (Group K), and the \textbf{flow} of specializations year over year decomposed into entry and exit (Group J). All four panels report the national mean computed across counties where the relevant variable is defined.

Panel A plots the within-entity count of skills at $\mathrm{RCA}>1$. The four entity types separate cleanly by scale. Corporate breadth rises from a mean of 416 specialized skills per county in 2010 to 943 in 2024, while university breadth tracks at roughly one-quarter to one-third of the corporate level (139 to 245), and federal-lab and government breadth occupy the lower band (91 to 122 and 71 to 167, respectively). The within-entity RCA is computed against each entity's own national skill pool, so the level differences reflect the breadth of skills each sector demands at scale, not the absolute size of its posting volume. All four series trend upward over the panel, consistent with the broader expansion of the Lightcast skill taxonomy and the trend toward longer, more skill-listed postings already documented at the aggregate level (\Cref{fig:mean_skills}).

Panel B plots the cosine similarity between each public-sector entity and corporate, the three policy-relevant ``versus corporate'' pairs. All three alignments rise modestly across the window. University-corporate alignment increases from 0.30 in 2010 to 0.42 in 2024; government-corporate alignment rises from 0.27 to 0.41; federal lab-corporate alignment is the flattest, edging from 0.33 to 0.36. Public-sector skill demand becomes more aligned with corporate skill demand over the panel, with the largest gain in the university channel.

Panels C and D decompose the Group J year-over-year flow into entry and exit shares. The entry share is the count of skills newly at $\mathrm{RCA}>1$ this year divided by the entity's current-year breadth; the exit share is the count of skills that lost $\mathrm{RCA}>1$ this year divided by the entity's prior-year breadth. Both shares lie in $[0,1]$ and are comparable across counties of widely different posting volumes, which the raw entry/exit counts are not. Each share reads as ``this entity's specialization mix turns over by $X$ percent a year.'' Two patterns are visible. First, the four entity types are ordered consistently across both panels: corporate has the most stable portfolio (entry share around 0.43, exit share around 0.43 in 2024), and the smaller public-sector entities reshuffle a larger fraction of their narrower portfolios each year (university entry 0.53, exit 0.58; federal lab entry 0.61, exit 0.70; government entry 0.55, exit 0.60 in 2024). Second, federal-lab and government exit shares trend upward over the panel (federal lab 0.56 to 0.70; government 0.54 to 0.60), while entry shares decline modestly in those sectors. The two public-research entities are losing a larger fraction of their prior-year specializations in recent years than they did in the early 2010s, even as their stocks in Panel A continue to expand with the taxonomy. The Group J flow and the Group I stock therefore capture orthogonal properties of the same entity-skill matrix.

\section{Usage Notes}
\label{sec:usage_notes}

This section consolidates guidance for users of the released panel. It covers
loading conventions, four known data caveats that constrain how specific
variables should be used, and the boundary conditions of the current release.

\subsection{Loading the panel}
\label{sec:usage_loading}

The panel is distributed as a single parquet file
(\texttt{county\_year\_panel.parquet}, zstd-compressed) and a mirror CSV
(\texttt{county\_year\_panel.csv}). The 5-digit \texttt{county} (FIPS) column
is a string; all other substantive columns are numeric. CSV loaders must
preserve leading zeros on \texttt{county} explicitly, since type-inference
utilities (e.g.\ \texttt{d3.autoType}, \texttt{pandas.read\_csv} default
behavior on integer-looking strings) will coerce \texttt{"06037"} to
\texttt{6037} and silently break county joins. Numeric NULLs must also be
preserved rather than mapped to zero; several variables documented below are
genuinely missing in early years rather than equal to zero. The codebook
(\texttt{data/codebook.md}) lists the 37 core columns (Groups A--H) sufficient
for most descriptive, teaching, and applied uses; the remaining 164 extension
columns (Groups I, J, K) are intended for sectoral and employer-pair analyses.

\subsection{The 2017--2018 vendor-methodology break}
\label{sec:usage_break}

Lightcast revised its posting-deduplication and source-coverage pipeline
between 2017 and 2018, producing a level shift in posting counts and a smaller
shift in skill-mention counts that is not attributable to underlying labor
demand. The break affects count-based variables in Groups B and C, the modality
counts in Group D, and any subsequent measure constructed from raw posting
totals. Three usable workarounds, in increasing order of robustness: (i) include county fixed effects and year fixed effects in any panel regression, in which case the level shift is absorbed by the year effects under the (testable) assumption that the methodology change is geographically uniform; (ii) prefer share-based and ratio-based variables (skill-type shares in Group~E, modality shares in Group~D for 2018 onward, RCA breadth in Group~F, complexity indices in Group~G) over raw count variables, since shares are first-order invariant to the level shift; and (iii) for studies focused on dynamics, restrict to 2018--2024 or estimate separate pre- and post-break trends.

\subsection{State-level placeholder FIPS codes (\texttt{XX999})}
\label{sec:usage_fips999}

Postings whose location resolves only to a state, rather than to a specific
county, are assigned a state-level placeholder FIPS code of the form
\texttt{XX999}, where \texttt{XX} is the 2-digit state FIPS prefix (for
example, \texttt{01999} for Alabama, \texttt{06999} for California,
\texttt{48999} for Texas). These placeholders are retained in the released
panel as standalone county-year rows; 765 such rows appear across 51
state-level codes, covering every state plus the District of Columbia.
Remote and hybrid postings are over-represented in these rows, since
employers often list remote postings against a state without a city or
county.

Users running strict county-level analyses should filter these rows via
\texttt{county.str.endswith("999")} or an equivalent test; users interested
in state-level coverage may keep them. The Connecticut planning-region remap
(8 historic counties to 9 planning regions, effective June 2022) is applied
uniformly across the panel; the new 9-region FIPS codes are used throughout
2010--2024, not just from 2022 onward.

\subsection{Fitness-complexity stability}
\label{sec:usage_fitness}

The Tacchella et~al.~\citep{tacchella2012new} fitness-complexity algorithm has
known numerical-stability issues at the tails: in very diversified counties
the iterated fitness score grows large, and in very specialized counties with
few $\mathrm{RCA}>1$ skills the harmonic-mean structure of the complexity
equation amplifies small denominators. The released \texttt{fitness} column
reports the log-transformed and 99th-percentile-winsorized value, computed
after 50 algorithm iterations and global standardization within year. Raw
fitness is not released. Users comparing fitness to the unwinsorized
literature should expect attenuation in the upper tail; users running OLS on
fitness should not re-transform.

\subsection{Low-volume noise in Groups I, J, and K}
\label{sec:usage_lowvol}

The 164 extension variables in Groups I (entity-type RCA breadth), J
(per-entity-type dynamics), and K (pairwise similarity between entity types) are
computed within entity-type-specific skill pools. Non-corporate entities, and government
and federal labs in particular, have postings concentrated in a smaller set of
county-years than corporate postings. Even when an entity has nonzero postings
in a county-year, thin volume destabilizes the entity-specific RCA denominator,
and the resulting breadth, similarity, and churn measures are noisy. The
codebook flags each Group I/J/K cell with the underlying entity-type posting
count; we recommend the following thresholds for downstream use: (i) Group I (breadth): entity-type posting count $\geq 100$ in the
county-year; (ii)  Group K (pairwise similarity): both entity types in
the pair at $\geq 100$ postings in the county-year; and (iii) Group J (dynamics): entity-type posting count $\geq 100$ in
both years of the year-over-year comparison.

\section*{Data Availability and Interactive Dashboard Access}
\label{sec:data_availability}

The county-year panel is deposited at \url{https://doi.org/10.5281/zenodo.20330086} and mirrored at the project GitHub repository \url{https://github.com/AntJam-Howell/skills-econ-geog-data}. The companion interactive dashboard is hosted at \url{https://skills-econ-geog.netlify.app/} and is open access. See \Cref{sec:records} for full details of file formats, variable groups, and folder structure.

\section*{Code Availability}
\label{sec:code}

Source code for replicating the construction of the released county-year panel is openly available at \url{https://github.com/AntJam-Howell/skills-econ-geog-data} under the MIT License. The pipeline is written in Python 3.11; exact dependency versions are pinned in \texttt{code/requirements.txt}. SLURM job scripts for the ASU Sol HPC environment are included in the repository. The pipeline is deterministic given the same Lightcast input shards and code version.

\section*{Funding}

This material is based upon work supported by the National Science Foundation under Grant No. 2431853 and the Alfred P. Sloan Foundation under Grant No. 2024-22666. Any opinions, findings, and conclusions or recommendations expressed in this material are those of the authors and do not necessarily reflect the views of the National Science Foundation or the Alfred P. Sloan Foundation.

\clearpage
\newpage

\clearpage
\appendix

\section*{Appendix}
\addcontentsline{toc}{section}{Appendix}

\renewcommand{\thetable}{A\arabic{table}}
\setcounter{table}{0}
\renewcommand{\thefigure}{A\arabic{figure}}
\setcounter{figure}{0}

\newpage

\begin{table}[!htbp] \centering
  \caption{Summary of released county-year variable groups.}
  \label{tab:variable_groups}
  \begin{threeparttable}
  \small
  \begin{tabular}{p{2.9cm} p{0.7cm} p{4.0cm} p{7.5cm}}
  \toprule
  \textbf{Dimension} & \textbf{Code} & \textbf{Group} & \textbf{Representative variables} \\
  \midrule
  \multicolumn{4}{l}{Core variables (37 variables)} \\
  \midrule
  Identifiers & A & Unit identifiers & County FIPS; calendar year (2 variables) \\
  \addlinespace[2pt]
  \multirow{2}{=}{Who is hiring}
  & B & Posting and skill-mention totals
  & Total postings; postings with at least one parsed skill; total mention counts of specialized, software, and common skills (5 variables) \\
  \addlinespace[2pt]
  & C & Posting counts by entity type
  & Postings by entity type: corporate (all private-sector), university, federal lab, and government (4 variables) \\
  \addlinespace[2pt]
  Nature of work
  & D & Work modality and internships
  & Posting counts and shares for remote, hybrid, and on-site; internship-flagged posting count (7 variables) \\
  \addlinespace[2pt]
  \multirow{4}{=}{Composition and structure of skill demand}
  & E & Skill composition
  & Shares of specialized, software, and common skills; mean skill mentions per posting; the percent of postings with at least one parsed skill (5 variables) \\
  \addlinespace[2pt]
  & F & Diversity, concentration, and complexity
  & Distinct-skill count; $\mathrm{RCA}>1$ breadth; average ubiquity; HHI; Shannon entropy; ECI \citep{hidalgo2009building}; Tacchella fitness-complexity \citep{tacchella2012new} (7 variables) \\
  \addlinespace[2pt]
  & G & Relatedness and network position
  & Average Balland skill density \citep{hidalgo2007product,balland2019smart}; Neffke skill coherence \citep{neffke2011how,boschma2017relatedness}; average network centrality of $\mathrm{RCA}>1$ skills (3 variables) \\
  \addlinespace[2pt]
  & H & Aggregate year-over-year dynamics
  & $\mathrm{RCA}>1$ entries, exits, and net counts; cosine distance between consecutive years' skill frequency vectors (4 variables) \\
  \midrule
  \multicolumn{4}{l}{Entity-decomposed extension (164 variables)} \\
  \midrule
  \multirow{2}{=}{Entity-specific measures}
  & I & Specialization breadth by entity type
  & $\mathrm{RCA}>1$ counts within each of the four entity-type-specific skill pools: corporate, university, federal lab, and government (4 variables) \\
  \addlinespace[2pt]
  & J & Per-employer-type dynamics
  & Each entity type's own $\mathrm{RCA}>1$ churning entries, exits, net, and cosine distance between consecutive years' entity-type-specific skill-frequency vectors (16 variables) \\
  \midrule
  Entity-pair measures
  & K & Employer-pair skill similarity
  & Six measure families (cosine, Jaccard, Hidalgo proximity \citep{hidalgo2007product}, weighted RCA overlap, directional gap count, directional gap relatedness) for six entity-type pairs (univ-corp, fede-corp, gove-corp, univ-fede, univ-gove, and fede-gove), each computed over all skills and separately over specialized, software, and common subsets (144 variables) \\
  \bottomrule
  \end{tabular}
  \begin{tablenotes}\footnotesize
  \item Notes: The 37 variables in groups A--H are the core measures of aggregate county-year labor and skill demand. The 164 variables in the entity-decomposed extension split into entity-specific measures (groups I and J) reporting each entity type's own specialization and dynamics, and entity-pair measures (group K) reporting alignment, overlap, and skill gaps between pairs of entity types within the same county-year.
  \end{tablenotes}
  \end{threeparttable}
\end{table}

\clearpage

\begin{table}[!htbp] \centering
  \caption{Summary statistics of core variables.}
  \label{tab:summary_stats}
  \begin{threeparttable}
  \scriptsize
  \setlength{\tabcolsep}{3pt}
  \begin{tabular}{lrrrrrrrrr}
  \toprule
  Variable & N & \%NA & Mean & SD & Min & P25 & Median & P75 & Max \\
  \midrule
\multicolumn{10}{l}{B. Labor demand: posting and skill-mention totals} \\
\midrule
\texttt{total\_postings} & 47,891 & 0.0 & 9,055 & 38,762 & 1 & 281 & 924 & 3,613 & 1,449,915 \\
\texttt{n\_has\_skill} & 47,891 & 0.0 & 8,589 & 37,248 & 0 & 250 & 841 & 3,337 & 1,384,092 \\
\texttt{mention\_specialized} & 47,891 & 0.0 & 67,498 & 331,264 & 0 & 1,354 & 4,795 & 20,060 & 11,593,285 \\
\texttt{mention\_software} & 47,891 & 0.0 & 10,080 & 60,815 & 0 & 67 & 289 & 1,520 & 1,941,841 \\
\texttt{mention\_common} & 47,891 & 0.0 & 42,575 & 202,561 & 0 & 886 & 3,183 & 13,568 & 7,551,771 \\
\midrule
\multicolumn{10}{l}{C. Labor demand: entity-type posting counts} \\
\midrule
\texttt{n\_corporate} & 47,891 & 0.0 & 8,600 & 37,071 & 0 & 262 & 864 & 3,382 & 1,387,229 \\
\texttt{n\_university} & 47,891 & 0.0 & 259 & 1,228 & 0 & 1 & 6 & 72 & 47,460 \\
\texttt{n\_federal\_lab} & 47,891 & 0.0 & 34 & 254 & 0 & 0 & 0 & 3 & 14,043 \\
\texttt{n\_government} & 47,891 & 0.0 & 162 & 644 & 0 & 4 & 19 & 82 & 37,450 \\
\midrule
\multicolumn{10}{l}{D. Nature of work: modality and internships} \\
\midrule
\texttt{n\_remote} & 47,891 & 0.0 & 317 & 2,141 & 0 & 2 & 11 & 54 & 106,608 \\
\texttt{n\_hybrid} & 47,891 & 0.0 & 439 & 3,204 & 0 & 1 & 6 & 51 & 196,667 \\
\texttt{n\_onsite} & 47,891 & 0.0 & 8,239 & 34,553 & 1 & 269 & 886 & 3,408 & 1,280,334 \\
\texttt{share\_remote} & 47,891 & 0.0 & 0.021 & 0.033 & 0 & 0.005 & 0.012 & 0.025 & 0.710 \\
\texttt{share\_hybrid} & 47,891 & 0.0 & 0.025 & 0.052 & 0 & 0.0004 & 0.005 & 0.017 & 0.833 \\
\texttt{share\_onsite} & 47,891 & 0.0 & 0.952 & 0.066 & 0.167 & 0.943 & 0.978 & 0.990 & 1.00 \\
\texttt{n\_internship} & 47,891 & 0.0 & 120 & 694 & 0 & 1 & 5 & 30 & 35,155 \\
\midrule
\multicolumn{10}{l}{E. Skill composition} \\
\midrule
\texttt{share\_specialized} & 47,891 & 0.0 & 0.576 & 0.054 & 0 & 0.548 & 0.570 & 0.597 & 1.00 \\
\texttt{share\_software} & 47,891 & 0.0 & 0.042 & 0.026 & 0 & 0.026 & 0.036 & 0.052 & 0.421 \\
\texttt{share\_common} & 47,891 & 0.0 & 0.382 & 0.056 & 0 & 0.355 & 0.386 & 0.413 & 1.00 \\
\texttt{mean\_skills\_per\_posting} & 47,891 & 0.0 & 9.46 & 3.01 & 0 & 7.63 & 9.29 & 11.0 & 157.0 \\
\texttt{pct\_has\_skill} & 47,891 & 0.0 & 90.9 & 6.32 & 0 & 88.7 & 92.4 & 95.0 & 100.0 \\
\midrule
\multicolumn{10}{l}{F. Skill diversity, concentration, and complexity} \\
\midrule
\texttt{n\_distinct\_skills} & 47,891 & 0.0 & 2,384 & 3,066 & 1 & 527 & 1,206 & 2,832 & 21,624 \\
\texttt{n\_rca\_skills} & 47,891 & 0.0 & 1,362 & 1,355 & 1 & 432 & 895 & 1,779 & 8,916 \\
\texttt{avg\_ubiquity} & 47,891 & 0.0 & 806.9 & 286.0 & 84.3 & 617.3 & 814.7 & 996.4 & 2,516 \\
\texttt{skill\_hhi} & 47,891 & 0.0 & 0.007 & 0.018 & 0.002 & 0.004 & 0.005 & 0.007 & 1.00 \\
\texttt{skill\_entropy} & 47,891 & 0.0 & 8.75 & 1.16 & 0 & 8.10 & 8.90 & 9.56 & 11.0 \\
\texttt{eci} & 47,891 & 0.0 & -0.0000 & 1.00 & -1.45 & -0.601 & -0.342 & 0.195 & 8.12 \\
\texttt{fitness} & 47,891 & 0.0 & 1.00 & 16.8 & 0.0000 & 0.0000 & 0.0000 & 0.0000 & 998.6 \\
\midrule
\multicolumn{10}{l}{G. Skill relatedness and network position} \\
\midrule
\texttt{skill\_density} & 47,891 & 0.0 & 0.052 & 0.058 & 0.0000 & 0.015 & 0.031 & 0.064 & 0.394 \\
\texttt{skill\_coherence} & 47,891 & 0.0 & 0.219 & 0.061 & 0.078 & 0.176 & 0.217 & 0.256 & 0.783 \\
\texttt{avg\_centrality} & 47,891 & 0.0 & 0.071 & 0.010 & 0.036 & 0.064 & 0.071 & 0.078 & 0.110 \\
\midrule
\multicolumn{10}{l}{H. Year-over-year dynamics} \\
\midrule
\texttt{churning\_entries} & 44,705 & 6.7 & 722 & 667 & 0 & 264 & 502 & 931 & 5,620 \\
\texttt{churning\_exits} & 44,705 & 6.7 & 664 & 630 & 0 & 230 & 453 & 859 & 6,068 \\
\texttt{churning\_net} & 44,705 & 6.7 & 58 & 269 & -4,017 & -47 & 38 & 142 & 4,527 \\
\texttt{skill\_cosine\_distance} & 44,685 & 6.7 & 0.143 & 0.160 & 0.0004 & 0.034 & 0.088 & 0.190 & 1.00 \\
  \bottomrule
  \end{tabular}
  \begin{tablenotes}\footnotesize
  \item Notes: Summary statistics for the 35 core variables in groups B--H; the two group-A identifiers (\texttt{county}, \texttt{year}), along with the within-county entity-based measures  (groups I, J, K) are omitted. 
  \end{tablenotes}
  \end{threeparttable}
\end{table}

\end{document}